\documentclass[pra,aps,twocolumn,superscriptaddress,amsmath,amssymb,floatfix,citeautoscript]{revtex4-2}
\usepackage{graphicx}
\usepackage{bm}
\usepackage{amsmath,amssymb}
\usepackage{xcolor}
\usepackage{bbold}    
\usepackage{todonotes} 
\usepackage[normalem]{ulem} 
\usepackage[nolist]{acronym}
\usepackage{orcidlink} 

\usepackage{amsfonts}
\usepackage{amsmath}
\usepackage{amssymb}
\usepackage{graphicx}

\newtheorem{proposition}{Proposition}
\newtheorem{corollary}{Corollary}

\def\be{ \begin{equation} }
\def\ee{ \end{equation} }
\def\bse{ \begin{subequations} }
\def\ese{ \end{subequations} }

\begin{document}

\title{Finite-Frequency Boson-Number Bounds Across the Quantum Rabi Transition}

\author{Razmik G. Unanyan}
\affiliation{Department of Physics and Research Center OPTIMAS, Rheinland-Pf\"alzische Technische Universit\"at Kaiserslautern-Landau, 67663 Kaiserslautern, Germany}
\author{Michael Fleischhauer}
\affiliation{Department of Physics and Research Center OPTIMAS, Rheinland-Pf\"alzische Technische Universit\"at Kaiserslautern-Landau, 67663 Kaiserslautern, Germany}
\author{Nikolay V. Vitanov}
\affiliation{Center for Quantum Technologies, Faculty of Physics, Sofia University ``St. Kliment Ohridski'', 5 James Bourchier Blvd., 1164 Sofia, Bulgaria}

\date{\today}

\begin{abstract}
We consider the quantum Rabi model, one of the most fundamental models of quantum optics, which describes a two-level system with transition frequency $\Omega$ coupled to a bosonic mode of frequency $\omega_0$ including counter-rotating terms. In the limit of large frequency separation $\Omega \gg \omega_0$, the model exhibits a quantum phase transition at some critical value of the coupling strength. While the model is exactly solvable,  no simple analytic expressions exist for important physical quantities. Here we derive rigorous bounds for the ground state energy and the boson occupation number across the phase transition for finite values of the frequency ratio $\Omega/\omega_0$.  
Exact diagonalization illustrates the tightness of the bounds in the subcritical, critical, and supercritical regimes, and we outline direct trapped-ion tests of the resulting finite-frequency inequalities.
\end{abstract}

\maketitle

This article is dedicated to the memory of Joseph H. Eberly, whose foundational insights into the quantum dynamics of two-level systems coupled to quantized radiation continue to inspire the field.

\section{Introduction}

The interaction of a two-level system with an oscillatory field is one of the oldest and most persistent paradigms of quantum physics, originating in Rabi's treatment of driven spin dynamics \cite{Rabi1936,Rabi1937}. When the field is itself quantized, the corresponding single-mode light-matter problem is described by the quantum Rabi model (QRM),
\begin{equation}
H=\omega_0 a^\dagger a+\frac{\Omega}{2}\sigma_z
+g\sigma_x(a^\dagger+a),
\label{eq:Rabi}
\end{equation}
where $\omega_0$ is the bosonic-mode frequency, $\Omega$ is the level splitting of the two-level system, and $g$ is the coupling strength. In the weak-coupling, near-resonant regime, neglecting the counter-rotating terms reduces Eq.~\eqref{eq:Rabi} to the famous Jaynes-Cummings model \cite{JaynesCummings1963}. Joe Eberly and co-workers established some of the most characteristic consequences of the latter model for a quantized single mode, notably the collapse and revival of atom--field coherence \cite{EberlyNarozhnySanchezMondragon1980,NarozhnySanchezMondragonEberly1981}. The full QRM, however, retains both rotating and counter-rotating processes and displays qualitatively different physics once the light-matter coupling is no longer perturbative. Despite the presence of counter-rotating terms the model is exactly solvable and its exact spectral structure and analytic representations have consequently attracted considerable interest \cite{Braak2011,Chen2012,Xie2017}.

This broader, non-perturbative parameter regime is no longer only of formal interest. Ultrastrong and deep-strong light-matter coupling have been reached or simulated in superconducting circuits, trapped-ion and cold atoms platforms, where the rotating-wave approximation breaks down and the ground state can contain a substantial virtual-boson population and strong spin-boson correlations \cite{Niemczyk2010,Casanova2010,Yoshihara2017,FornDiaz2019,Pedernales2015,Lv2018, weitz2023,Weitz2024}. Furthermore a realization in an effective Raman-coupled cavity system was suggested in \cite{dimer2007proposed}.
The resulting departure from Jaynes--Cummings physics, including the qubit--oscillator dynamics in the ultrastrong-coupling and quasi-degenerate regimes, was analyzed by Agarwal, Hashemi Rafsanjani, and Eberly \cite{AgarwalRafsanjaniEberly2013}. These developments make ground-state observables of the full QRM, and in particular the boson occupation, directly relevant to experimentally accessible regimes rather than merely asymptotic theoretical quantities.

A further remarkable feature of the QRM is that it supports quantum critical behavior despite containing only one two-level system and one bosonic mode. Phase-transition-like behavior in the low-oscillator-frequency limit was identified by Ashhab \cite{Ashhab2013}. Hwang, Puebla, and Plenio subsequently established the quantum phase transition and its universal finite-frequency scaling by identifying the singular large-frequency-ratio limit $\Omega/\omega_0\to\infty$ as the effective thermodynamic limit \cite{Hwang2015}. We introduce the parameters
\begin{equation}
\eta\equiv\frac{\Omega}{\omega_0},
\quad
\gamma\equiv\frac{2g}{\sqrt{\omega_0\Omega}},
\quad
N_g\equiv\langle\Phi_g|a^\dagger a|\Phi_g\rangle,
\label{eq:definitions}
\end{equation}
where $|\Phi_g\rangle$ is the exact ground state. In the limit $\eta\to\infty$, the ground state undergoes a continuous quantum phase transition at $\gamma=1$, or equivalently at $g_c=\sqrt{\omega_0\Omega}/2$ \cite{Hwang2015}. The critical structure also extends to excited states \cite{Puebla2016}. A trapped-ion protocol for probing the equilibrium and nonequilibrium critical scaling was developed in Ref.~\cite{Puebla2017}, and signatures of the ground-state transition were subsequently observed experimentally with a single trapped ion by measuring both the spin population and the phonon occupation \cite{Cai2021}.

The boson number provides a particularly transparent characterization of this transition. In the thermodynamic-limit $\eta\to\infty$ one finds 
\begin{equation}
\lim_{\eta\to\infty}\frac{N_g}{\eta}
=
\begin{cases}
0, & \gamma\leq 1,\\[1ex]
\dfrac{1}{4}\left(\gamma^2-\gamma^{-2}\right), & \gamma>1,
\end{cases}
\label{eq:intro_scaledlimit}
\end{equation}
and at the critical point the finite-frequency scaling is \cite{Hwang2015}
\begin{equation}
    N_g(\gamma=1) \propto\eta^{1/3}.
\end{equation}
Thus the three regimes are distinguished by parametrically different occupation scales: $O(1)$ for fixed $\gamma<1$, $O(\eta^{1/3})$ at $\gamma=1$, and $O(\eta)$ for fixed $\gamma>1$. At finite $\eta$, however, the exact ground state retains the $\mathbb{Z}_2$ parity symmetry of the QRM, so the extensive occupation above threshold should not be interpreted as a nonzero finite-$\eta$ expectation value $\langle a\rangle_g$. Rather, $N_g$ is the parity-even observable that records the growing bosonic displacement scale.

The limiting phase diagram does not  answer how accurately a finite-frequency system is described by these asymptotic results in the thermodynamic limit $\eta = \Omega/\omega_0 \to \infty$. This distinction is important both conceptually and practically: experiments always operate at finite $\eta$, and numerical diagonalization requires a finite oscillator cutoff whose necessary size grows with the boson occupation. Ground-state photon populations at finite parameters have been analyzed using variational, polaronic, and effective-Hamiltonian methods \cite{Liu2015,Ying2015,Gartner2022}, while exact formulations of the spectrum provide non-perturbative benchmarks \cite{Braak2011,Chen2012}. What is still desirable is a simple analytic control of the exact finite-$\eta$ occupation that is uniform enough to distinguish the normal, critical, and superradiant scales and that quantifies the approach to the thermodynamic result.

There is also an important literature on rigorous results for spin-boson ground states. General non-perturbative results on existence and uniqueness were established by Arai and Hirokawa \cite{AraiHirokawa1997}. More directly relevant here, Hirokawa, M{\o}ller, and Sasaki derived rigorous dressed-photon estimates for a generalized QRM containing an $A^2$ term and developed the representation used to control the ground-state boson population \cite{HirokawaMollerSasaki2017}. Specialized to the ordinary QRM, this machinery yields the global upper estimate $N_g\leq g^2/\omega_0^2$. Such results provide rigorous non-perturbative control, but they are not tailored to the finite-frequency critical scaling problem: by themselves they neither confine $N_g$ to a relatively shrinking interval about the thermodynamic occupation for fixed $\gamma>1$ nor establish matching two-sided  bounds to $\eta^{1/3}$ at $\gamma=1$.

The goal of this work is to supply this finite-$\eta$ control. We first derive, from coherent-state upper and lower bounds, a ground-state-energy interval of width $\omega_0$. Combining this bracket with an operator inequality produces explicit upper and lower bounds on the ground-state photon occuopation number $N_g$. For every fixed $\gamma>1$ we show that
\begin{equation}
\left|N_g-N_{\rm cl}\right|
\leq\gamma\sqrt{\frac{\eta}{2}},
\quad
N_{\rm cl}=\frac{\eta}{4}\left(\gamma^2-\gamma^{-2}\right),
\label{eq:intro_mainbound}
\end{equation}
so the scaled occupation converges to its thermodynamic value $N_\textrm{cl}$ with an absolute error $O(\eta^{-1/2})$; equivalently, for every fixed $\gamma>1$, the relative error in $N_g$ is $O(\eta^{-1/2})$.
At $\gamma=1$, where the quadratic confinement of the soft mode vanishes, retaining the leading quartic term and using an explicit adiabatic variational state gives matching lower and upper bounds proportional to $\eta^{1/3}$. 
Hence $N_g=O(\eta^{1/3})$, meaning that it is bounded both above and below by positive constant multiples of $\eta^{1/3}$ for all sufficiently large $\eta$.
For fixed $\gamma<1$ the bounds establish $N_g=O(1)$. We additionally obtain a weak-coupling lower bound that is asymptotically sharp and a rigorous Fock-space tail estimate that converts the occupation ceiling into a sufficient numerical truncation criterion. In this way a single bounding framework resolves the  boson-number scales across the Rabi transition.

The  paper is organized as follows. Section~\ref{sec:energy} establishes the coherent-state energy bracket, reviews the normal-phase effective description, required to interpret the finite occupation below threshold, and derives complementary upper and lower resolvent bounds. Section~\ref{sec:bounds} combines these ingredients with an operator inequality to obtain the explicit finite-$\eta$ interval for $N_g$, then treats the critical point separately by retaining the quartic soft mode and derives the intensive thermodynamic limit. Section~\ref{sec:numerics} tests the bounds by exact diagonalization and converts the occupation ceiling into a rigorous Fock-space tail criterion. Section~\ref{sec:experiment} translates the dimensionless parameters to a trapped-ion implementation and describes direct experimental tests of the finite-frequency interval, the critical scaling, and the superradiant error bound. Finally, Sec.~\ref{sec:conclusions} summarizes  the results and their relation to earlier asymptotic and rigorous analyses.

\section{Preliminary bounds}
\label{sec:energy}

Before we derive new bounds for ground-state energy and boson number of the QRM in the following section, we will here present some known results based on coherent-state representations of the Hamiltonian \eqref{eq:Rabi}.

\paragraph{Ground-state energy bracket and coherent-state representations --}
Oscillator coherent states $\vert \alpha \rangle$ obey
\begin{equation}
\frac{1}{\pi}\int d^2\alpha\,|\alpha\rangle\langle\alpha|=\mathbf{1},
\label{eq:coherentresolution}
\end{equation}
 and the Glauber-$P$ representations of photon operators read
\begin{equation}
a=\frac{1}{\pi}\int d^2\alpha\,\alpha|\alpha\rangle\langle\alpha|,
\quad
a^\dagger=\frac{1}{\pi}\int d^2\alpha\,\alpha^*|\alpha\rangle\langle\alpha|.
\end{equation}
With this
\begin{equation}
a^\dagger a=aa^\dagger-1
=\frac{1}{\pi}\int d^2\alpha\,(|\alpha|^2-1)|\alpha\rangle\langle\alpha|.
\label{eq:numberP}
\end{equation}
and the (operator-valued)  $P$-representation of $H$ reads
\begin{equation}
P(\alpha)=\omega_0(|\alpha|^2-1)
+\frac{\Omega}{2}\sigma_z
+g(\alpha+\alpha^*)\sigma_x,
\label{eq:Psymbol}
\end{equation}
whereas the  Husimi
representation is given by
\begin{equation}
Q(\alpha)=\langle\alpha|H|\alpha\rangle
=\omega_0|\alpha|^2
+\frac{\Omega}{2}\sigma_z
+g(\alpha+\alpha^*)\sigma_x.
\label{eq:Qsymbol}
\end{equation}
In particular,
\begin{equation}
P(\alpha)=Q(\alpha)-\omega_0\mathbf{1}_\sigma.
\label{eq:PQshift}
\end{equation}
The two eigenvalues of $Q(\alpha)$ are
\begin{equation}
W_\pm(\alpha)=\omega_0|\alpha|^2
\pm\sqrt{\left(\frac{\Omega}{2}\right)^2
+4g^2(\operatorname{Re}\alpha)^2}.
\label{eq:Wpm}
\end{equation}
Minimization over $\operatorname{Im}\alpha$ first gives $\operatorname{Im}\alpha=0$. For $\gamma\leq1$ the remaining minimum of $W_-$ is at $\alpha=0$ and equals $-\Omega/2$. For $\gamma>1$ there are two degenerate minima with
\begin{equation}
(\operatorname{Re}\alpha_0)^2
=\frac{g^2}{\omega_0^2}-\frac{\Omega^2}{16g^2},
\end{equation}
and
\begin{equation}
\mathcal E(\gamma)\equiv\min_\alpha W_-(\alpha)=
\begin{cases}
-\dfrac{\Omega}{2}, & \gamma\leq1,\\[1ex]
-\dfrac{\Omega}{4}\left(\gamma^2+\gamma^{-2}\right), & \gamma>1.
\end{cases}
\label{eq:Eclass}
\end{equation}

The $P$ and $Q$ representations
 give directly a bracket to the
ground-state energy.  Since the lowest eigenvalue of $Q(\alpha)$ is $W_-(\alpha)\geq\mathcal E(\gamma)$, Eq.~\eqref{eq:PQshift} implies the spin-space operator inequality
\begin{equation}
P(\alpha)\geq
\left[\mathcal E(\gamma)-\omega_0\right]\mathbf{1}_\sigma
\quad\text{for every }\alpha.
\label{eq:Ppointwise}
\end{equation}
Using the $P$ representation
\begin{equation}
H=\frac{1}{\pi}\int d^2\alpha\,
|\alpha\rangle\langle\alpha|\otimes P(\alpha)
\label{eq:Hupperrepresentation}
\end{equation}
together with Eq.~\eqref{eq:coherentresolution} therefore gives
\begin{equation}
H\geq
\left[\mathcal E(\gamma)-\omega_0\right]\mathbf{1},
\end{equation}
and hence $E_g\geq\mathcal E(\gamma)-\omega_0$. Conversely, let $|\chi_-(\alpha_0)\rangle$ be the normalized spin eigenstate associated with the lower eigenvalue of $Q(\alpha_0)$ at a minimizing point $\alpha_0$. The coherent-product variational state $|\alpha_0\rangle\otimes|\chi_-(\alpha_0)\rangle$ has expectation value $\mathcal E(\gamma)$, so the variational principle gives $E_g\leq\mathcal E(\gamma)$. Thus
\begin{equation}
\mathcal E(\gamma)-\omega_0\leq E_g\leq\mathcal E(\gamma).
\label{eq:Ebounds}
\end{equation}
This direct zero-temperature argument is the operator/variational content of the corresponding Berezin--Lieb coherent-state construction \cite{Lieb1973,Berezin1975} and avoids introducing a finite-temperature partition function.

The constructed energy interval has width $\omega_0$. It becomes absolutely sharp when $\omega_0\to0$ at fixed $\Omega$ and $\gamma$; this statement does not assert, however, that no narrower  bracket is possible for finite $\omega_0$. 
If instead $\Omega\to\infty$ at fixed $\omega_0$, its absolute width remains $\omega_0$ but becomes negligible relative to the $O(\Omega)$ ground-state energy. In either realization of the singular limit, Eq.~\eqref{eq:Eclass} has a continuous first derivative and a discontinuous second derivative at $\gamma=1$, recovering the known second-order transition \cite{Hwang2015,Puebla2016}.

\paragraph{Normal phase and effective potential --}
The leading energy $\mathcal E=-\Omega/2$ for $\gamma<1$ must not be differentiated by itself to infer $N_g=0$. The reason is that Eq.~\eqref{eq:Ebounds} leaves an $O(\omega_0)$ uncertainty, and an $O(\omega_0)$ correction to $E_g$ gives an $O(1)$ contribution to $\partial E_g/\partial\omega_0$. The Hellmann-Feynman identity
\begin{equation}
N_g=\left.\frac{\partial E_g}{\partial\omega_0}\right|_{g,\Omega}
\label{eq:HF}
\end{equation}
is therefore sensitive precisely to the part of the energy not fixed by Eq.~\eqref{eq:Ebounds}.

The standard low-energy expansion makes the normal-phase physics explicit \cite{Hwang2015}. To leading nontrivial order in $1/\Omega$,
\begin{equation}
H_{\rm eff}=-\frac{\Omega}{2}+\omega_0a^\dagger a
-\frac{g^2}{\Omega}(a+a^\dagger)^2+\cdots.
\label{eq:Heffnormal}
\end{equation}
For $\gamma<1$ this is a stable squeezed harmonic oscillator with frequency
\begin{equation}
\omega_{\rm eff}=\omega_0\sqrt{1-\gamma^2}.
\label{eq:omegaeff}
\end{equation} 
Its ground-state energy and boson number are
\begin{align}
E_g&=-\frac{\Omega}{2}
+\frac{\omega_0}{2}\left(\sqrt{1-\gamma^2}-1\right)+o(\omega_0),
\label{eq:Enormal}\\
N_{\rm normal}&=\frac{1}{4}\left[
\frac{1}{\sqrt{1-\gamma^2}}+\sqrt{1-\gamma^2}-2\right]+o(1).
\label{eq:Nnormal}
\end{align}
Thus $N_g$ is finite and generally nonzero for every fixed $0<\gamma<1$. The normal phase is characterized by the absence of an extensive displacement, not by an exactly empty oscillator.

The same conclusion follows from an adiabatic-potential picture once the oscillator quadratures are normalized consistently. Define $x=(a+a^\dagger)/\sqrt2$ and $p=(a-a^\dagger)/(i\sqrt2)$, so that $[x,p]=i$ and $a^\dagger a=(x^2+p^2-1)/2$. After placing the fast spin (Born-Oppenheimer approximation) in its lower eigenstate, the coordinate-dependent part of the lower adiabatic surface is
\begin{equation}
V_-(x)=\frac{\omega_0}{2}x^2
-\sqrt{\left(\frac{\Omega}{2}\right)^2+2g^2x^2},
\label{eq:Veff}
\end{equation}
with the kinetic term $\omega_0p^2/2$ and the constant $-\omega_0/2$ omitted. Near the origin,
\begin{equation}
V_-(x)=-\frac{\Omega}{2}
+\frac{\omega_0}{2}(1-\gamma^2)x^2+O(x^4).
\label{eq:Vexpand}
\end{equation}
The curvature changes sign exactly at $4g^2=\omega_0\Omega$, i.e. $\gamma=1$. For $\gamma<1$ the state is centered at $x=0$ but remains squeezed.

\paragraph{Supercritical phase --}
The effective potential can also be used to discuss the supercritical phase. For
$\gamma>1$ its lower surface develops two displaced minima. In terms of the coherent-state coordinate used above, their squared displacement is
\begin{equation}
|\alpha_0|^2=\frac{g^2}{\omega_0^2}-\frac{\Omega^2}{16g^2}
=\frac{\eta}{4}(\gamma^2-\gamma^{-2})\equiv N_{\rm cl}.
\label{eq:Ncl}
\end{equation}
At any finite $\eta$, however, the Rabi Hamiltonian has the parity symmetry $\Pi=\sigma_z(-1)^{a^\dagger a}$, and its nondegenerate ground state can be chosen with definite parity. Hence $\langle a\rangle_g=0$ even in the regime where the adiabatic surface has minima at $\pm\alpha_0$. The macroscopic quantity controlled below is therefore the parity-even occupation $N_g$, not a finite-$\eta$ symmetry-breaking expectation value. A nonzero $\langle a\rangle$ arises only after selecting one of the degenerate symmetry-broken branches in the singular limit.

\paragraph{Resolvent bounds --}
Closely related identities and photon-number bounds were used in rigorous spin-boson analyses and, for the generalized Rabi model, explicitly by Hirokawa, M{\o}ller, and Sasaki \cite{AraiHirokawa1997,HirokawaMollerSasaki2017}. We include the short derivation because the resulting bound enters our final finite-frequency interval.
From $[H,a]=-\omega_0a-g\sigma_x$ and $H|\Phi_g\rangle=E_g|\Phi_g\rangle$,
 one obtains
\be
a|\Phi_g\rangle=-g(H-E_g+\omega_0)^{-1}\sigma_x|\Phi_g\rangle.
\ee
Since $H-E_g\geq0$, we find
\begin{align}
N_g&=g^2\langle\Phi_g|\sigma_x
(H-E_g+\omega_0)^{-2}
\sigma_x|\Phi_g\rangle\nonumber\\
&\leq\frac{g^2}{\omega_0^2}
=\frac{\gamma^2\eta}{4}.
\label{eq:resolventbound}
\end{align}
For the ordinary Rabi model this is the $A^2\to0$ specialization of the earlier dressed-photon upper estimate in Ref.~\cite{HirokawaMollerSasaki2017}. The squared resolvent is essential.

A further simple consequence of the same identity is a parameter-only lower bound. Let $A=H-E_g+\omega_0>0$ and $|\chi\rangle=\sigma_x|\Phi_g\rangle$. Since $f(x)=x^{-2}$ is convex on the positive axis, Jensen's inequality for the spectral measure of $A$ \cite{Hansen2003} gives
\begin{equation}
\langle\chi|A^{-2}|\chi\rangle
\geq
\langle\chi|A|\chi\rangle^{-2}.
\label{eq:jensenresolvent}
\end{equation}
Moreover,
\begin{equation}
\sigma_xH\sigma_x=H-\Omega\sigma_z,
\quad
\langle\chi|A|\chi\rangle
=\omega_0-\Omega\langle\sigma_z\rangle_g
\leq\omega_0+\Omega.
\label{eq:sigmaxconj}
\end{equation}
Therefore
\begin{equation}
N_g\geq N_{\rm L}^{(R)}\equiv\frac{g^2}{(\Omega+\omega_0)^2}
=\frac{\gamma^2\eta}{4(1+\eta)^2}.
\label{eq:resolventlower}
\end{equation}
This parameter-only bound is nonzero for every $g\neq0$. It is also asymptotically sharp in weak coupling: ordinary perturbation theory gives
\begin{equation}
N_g=\frac{g^2}{(\Omega+\omega_0)^2}+O(g^4),
\quad g\to0.
\label{eq:weakcouplingN}
\end{equation} 
Thus the two resolvent estimates control opposite limits particularly well: the lower bound is tight at weak coupling, while the upper bound approaches the leading deep-strong-coupling occupation.

\section{Boson-number bounds and critical asymptotics}
\label{sec:bounds}

In the following we want to derive some better bounds both on the ground-state photon number and the ground-state energy away from the thermodynamic limit, i.e. for finite values of $\omega_0$.

The hamiltonian eq.\eqref{eq:Rabi} can be written as 
\begin{equation}
H=\omega_0a^\dagger a+B,\qquad
B=\frac{\Omega}{2}\sigma_z+g\sigma_x(a+a^\dagger).
\label{eq:Bdef}
\end{equation}
Since
\begin{equation}
B^2=\left(\frac{\Omega}{2}\right)^2
+g^2(a+a^\dagger)^2
\label{eq:Bsq}
\end{equation}
and every self-adjoint operator obeys $B\geq-|B|$, one has the operator inequality
\begin{equation}
H\geq \omega_0a^\dagger a
-\sqrt{\left(\frac{\Omega}{2}\right)^2+g^2(a+a^\dagger)^2}.
\label{eq:operatorineq}
\end{equation}
Taking the expectation value in the ground state $|\Phi_g\rangle$ and using the concavity of the square root gives
\begin{equation}
E_g\geq\omega_0N_g
-\sqrt{\left(\frac{\Omega}{2}\right)^2
+g^2\langle(a+a^\dagger)^2\rangle_g}.
\label{eq:concavity}
\end{equation}
Finally,
\begin{equation}
0\le -\frac{1}{4}(a^\dagger -a)^2 = a^\dagger a-\frac14(a+a^\dagger)^2+\frac12
\label{eq:quadratureopineq}
\end{equation}
implies
\begin{equation}
\langle(a+a^\dagger)^2\rangle_g
\leq4\left(N_g+\frac12\right),
\label{eq:quadraturemoment}
\end{equation}
so that
\begin{equation}
E_g\geq\omega_0N_g
-\sqrt{\left(\frac{\Omega}{2}\right)^2
+4g^2\left(N_g+\frac12\right)}.
\label{eq:keyNineq}
\end{equation}
Combining Eq.~\eqref{eq:keyNineq} with the upper energy bound $E_g\leq\mathcal E(\gamma)$ in Eq.~\eqref{eq:Ebounds} gives explicit restrictions on $N_g$.

\begin{proposition}

For $\eta=\Omega/\omega_0>0$ and $\gamma\geq0$, define
\begin{equation}
N_{\rm L}^{(E)}=
\begin{cases}
0, & \gamma\leq1,\\[1ex]
\displaystyle\max\!\left\{0,
\frac{\eta}{4}(\gamma^2-\gamma^{-2})
-\gamma\sqrt{\frac{\eta}{2}}\right\}, & \gamma>1,
\end{cases}
\label{eq:NL}
\end{equation}
and
\begin{equation}
N_{\rm U}^{(E)}=
\begin{cases}
\displaystyle\frac{\eta}{2}\left[
\gamma^2-1+
\sqrt{(\gamma^2-1)^2+\frac{2\gamma^2}{\eta}}
\right], & \gamma\leq1,\\[2ex]
\displaystyle\frac{\eta}{4}(\gamma^2-\gamma^{-2})
+\gamma\sqrt{\frac{\eta}{2}}, & \gamma>1.
\end{cases}
\label{eq:NUE}
\end{equation}
Then the ground-state occupation satisfies
\begin{equation}
\begin{aligned}
N_{\rm L}\equiv\max\{N_{\rm L}^{(E)},N_{\rm L}^{(R)}\}\leq N_g\leq N_{\rm U}\equiv
\min\left\{N_{\rm U}^{(E)},\frac{\gamma^2\eta}{4}\right\}.
\end{aligned}
\label{eq:finalbounds}
\end{equation}
\end{proposition}

For completeness, the root selection follows directly from Eq.~\eqref{eq:keyNineq}. For $\gamma\leq1$ one obtains
\be
N_g^2+\eta(1-\gamma^2)N_g-\gamma^2\eta/2\leq0,
\ee
 whose only nonnegative restriction is the upper root in Eq.~\eqref{eq:NUE}. For $\gamma>1$ the inequality is equivalent to
\begin{equation}
\left[N_g-\frac{\eta}{4}(\gamma^2-\gamma^{-2})\right]^2
\leq\frac{\gamma^2\eta}{2},
\label{eq:superquad}
\end{equation}
which gives the two roots in Eqs.~\eqref{eq:NL} and \eqref{eq:NUE}; when the lower root is negative, positivity of $N_g$ is the stronger energy-derived lower bound. The final interval in Eq.~\eqref{eq:finalbounds} combines these roots with the independent resolvent bounds \eqref{eq:resolventlower} and \eqref{eq:resolventbound}.

\paragraph{Asymptotic regimes.}

Equations~\eqref{eq:NL}-\eqref{eq:finalbounds} separate naturally into the three critical regimes.

For fixed $\gamma<1$ and $\eta\to\infty$,
\begin{equation}
N_{\rm U}^{(E)}
=\frac{\gamma^2}{2(1-\gamma^2)}+O(\eta^{-1}).
\label{eq:normalboundlimit}
\end{equation}
Thus the rigorous bounds prove $N_g=O(1)$, but they do not determine its exact limiting value. The squeezed-oscillator result in Eq.~\eqref{eq:Nnormal} supplies the corresponding asymptotic benchmark. The resolvent lower bound remains positive but vanishes as $N_{\rm L}^{(R)}\sim\gamma^2/(4\eta)$ at fixed $\gamma$.

At $\gamma=1$ the simpler estimate above gives only $N_g\leq\sqrt{\eta/2}$. The loss of sharpness has a clear origin: the quadratic confinement vanishes at the transition, and the leading restoring term is quartic. Keeping this term allows the critical exponent itself to be bounded rigorously.

With the quadratures $x$ and $p$ introduced above, the Hamiltonian at $\gamma=1$ can be written as
\begin{equation}
H=\frac{\omega_0}{2}(p^2+x^2-1)
+\frac{\Omega}{2}\sigma_z
+\sqrt2\,g x\sigma_x,
\quad
g^2=\frac{\Omega\omega_0}{4},
\label{eq:Hcritical}
\end{equation} 
For fixed $x$, let $|\chi_-(x)\rangle = \cos\theta(x) \vert \uparrow\rangle + \sin\theta(x)\vert \downarrow\rangle $ denote the lower eigenstate of the spin part and choose it real. Its mixing angle satisfies
\begin{equation}
\tan\theta(x)=\sqrt{\frac{2}{\eta}}\,x,
\label{eq:criticalmixing}
\end{equation}
With the real gauge choice $\langle\chi_-(x)|\partial_x\chi_-(x)\rangle=0$, differentiation of the mixing angle gives the Born--Huang term
\begin{equation}
\|\partial_x\chi_-(x)\|^2=
\frac{1}{2\eta(1+2x^2/\eta)^2}.
\label{eq:BHterm}
\end{equation}
For a normalized real $\phi\in H^1(\mathbb{R})$ with $x^4\phi\in L^2(\mathbb{R})$, the spinor wavefunction $|\Psi(x)\rangle=\phi(x)|\chi_-(x)\rangle$ belongs to the quadratic-form domain of $H$ and is therefore an admissible variational state. Acting with $p=-i\partial_x$ on both factors, $\phi(x)$ and $\vert \chi_-(x)\rangle$ gives
\begin{equation}
\langle p^2\rangle_\Psi
=
{\int_{\mathbb R}dx\,|\partial_x\phi(x)|^2
\
+ |\phi(x)|^2
\|\partial_x\chi_-(x)\|^2.}
\label{eq:kineticdecomposition}
\end{equation}
The cross term vanishes because $\phi$ is real and the chosen real normalized spinor satisfies $\langle\chi_-|\partial_x\chi_-\rangle=0$. Using $\sqrt{1+y}\geq1+y/2-y^2/8$ for $y\geq0$, together with $\|\partial_x\chi_-\|^2\leq1/(2\eta)$, yields
\begin{equation}
E_g\leq -\frac{\Omega}{2}-\frac{\omega_0}{2}
+\omega_0\left[\frac12\langle p^2\rangle_\phi
+\frac{\langle x^4\rangle_\phi}{4\eta}+\frac{1}{4\eta}\right],
\label{eq:criticalEvar}
\end{equation}
where we have introduced the short-hand notation:
\begin{equation}
    \langle a \rangle_\phi = \int\!\! dx\,  \phi(x) \, a \, \phi(x).
\end{equation}
(Note that $\phi(x)$ is real).
Choosing a minimum-uncertainty Gaussian $\phi(x)$ with
\begin{equation}
\langle x^2\rangle_\phi=v,
\quad
\langle p^2\rangle_\phi=\frac{1}{4v},
\quad
\langle x^4\rangle_\phi=3v^2,
\label{eq:gaussianmoments}
\end{equation}
and minimizing over $v$ gives
\begin{equation}
v=\left(\frac{\eta}{12}\right)^{1/3},
\label{eq:vopt}
\end{equation}
and
\begin{equation}
E_g\leq -\frac{\Omega}{2}-\frac{\omega_0}{2}+\omega_0 A_\eta,
\quad
A_\eta=c_E\eta^{-1/3}+\frac{1}{4\eta},
\label{eq:Aeta}
\end{equation}
where $c_E=\frac14(3/2)^{4/3}$.
Alternatively, one may use the ground-state energy of the quartic oscillator
\be
\frac{1}{2}p^2+\frac{x^4}{4\eta},
\ee
whose lowest eigenvalue is
\be
0.420805\,\eta^{-1/3}<c_E\eta^{-1/3}.
\ee
This numerical replacement improves the constant but is not used in the analytic bounds below.

Inequality \eqref{eq:keyNineq} for the occupation number also simplifies at criticality. Writing $s=N_g+1/2$, it is exactly
\begin{equation}
E_g\geq -\frac{\Omega}{2}-\frac{\omega_0}{2}
+\frac{\omega_0\eta}{4}
\left(\sqrt{1+\frac{4s}{\eta}}-1\right)^2.
\label{eq:criticalElower}
\end{equation}
Combining Eqs.~\eqref{eq:Aeta} and \eqref{eq:criticalElower} gives
\begin{equation}
s\leq\sqrt{\eta A_\eta}+A_\eta.
\label{eq:critical_s_upper}
\end{equation}
A matching lower scaling follows from the operator inequality
\begin{equation}
H\geq-\frac{\Omega}{2}-\frac{\omega_0}{2}
+\frac{\omega_0}{2}p^2,
\label{eq:critical_H_p}
\end{equation}
which in fact holds for every $\gamma\leq 1$. Taking its ground-state expectation value and comparing it with Eq.~\eqref{eq:Aeta} gives
\begin{equation}
\langle p^2\rangle_g\leq2A_\eta.
\label{eq:critical_p_upper}
\end{equation}
The Heisenberg relation $\langle x^2\rangle_g\langle p^2\rangle_g\geq1/4$ (the parity eigenstate has vanishing first moments) then implies
\begin{equation}
\langle x^2\rangle_g\geq\frac{1}{8A_\eta}.
\label{eq:critical_x_lower}
\end{equation}

\begin{proposition}
At $\gamma=1$, define $A_\eta$ by Eq.~\eqref{eq:Aeta}. Then
\begin{equation}
\begin{aligned}
N_g&\geq\max\!\left\{0,\frac{1}{16A_\eta}-\frac12,
\frac{\eta}{4(1+\eta)^2}\right\},\\
N_g&\leq\min\!\left\{\sqrt{\eta A_\eta}+A_\eta-\frac12,
\sqrt{\frac{\eta}{2}},\frac{\eta}{4}\right\}.
\end{aligned}
\label{eq:criticalbound}
\end{equation}
\end{proposition}
The first lower entry can be sharpened using the auxiliary-parameter inequality
\be
a^{\dagger}a-\frac{\lambda}{4}\bigl(a^{\dagger}+a\bigr)^2
\geq
\frac{1}{2}\bigl(\sqrt{1-\lambda}-1\bigr),
\ee
valid for every $0\leq\lambda\leq1$. This is the lower spectral edge of the quadratic oscillator on the left. Taking its ground-state expectation value and using $\langle(a+a^\dagger)^2\rangle_g=2\langle x^2\rangle_g\geq1/(4A_\eta)$ gives
\be
N_g\geq
\max\left\{
0,
\frac{1}{16A_{\eta}}+A_{\eta}-\frac{1}{2},
\frac{\eta}{4(1+\eta)^2}
\right\}.
\ee
%
For $A_\eta\leq1/4$, optimization gives $\lambda_*=1-16A_\eta^2$ and hence $N_g\geq1/(16A_\eta)+A_\eta-1/2$; for $A_\eta>1/4$, the admissible optimum is $\lambda_*=0$ and gives the trivial bound $N_g\geq0$. Thus the \textit{refined lower bound} is 
\begin{equation}
    N_g\geq\max\left\{B_\eta,\frac{\eta}{4(1+\eta)^2}\right\}
\end{equation}
with 
\begin{equation}
B_\eta=
\begin{cases}
\dfrac{1}{16A_\eta}+A_\eta-\dfrac12,&A_\eta\leq\dfrac14,\\[1ex]
0,&A_\eta>\dfrac14,
\end{cases}
\label{eq:Beta}
\end{equation}
This refinement changes no exponent but yields the constant stated below.
In particular,
\begin{equation}
c_L\eta^{1/3}-\frac12+O(\eta^{-1/3})
\leq N_g\leq
c_U\eta^{1/3}-\frac12+O(\eta^{-1/3}),
\label{eq:criticaltheta}
\end{equation}
with $c_L=\frac14(2/3)^{4/3}$ and $c_U=\frac12(3/2)^{2/3}$. Hence $N_g(\gamma=1)={\cal O}(\eta^{1/3})$.

The finite-frequency analysis of Ref.~\cite{Hwang2015}, based on a
Gaussian variational treatment of the critical quartic Hamiltonian,
{gave}
\begin{equation}
N_g(\gamma=1)\sim C_{\rm c}^{(\mathrm G)}\eta^{1/3},
\qquad
C_{\rm c}^{(\mathrm G)}
=\frac16\left(\frac32\right)^{2/3}.
\label{eq:knowncritical}
\end{equation}
This result correctly predicts the critical exponent, but its
prefactor is variational  {and not } exact. For the quartic scaling
Hamiltonian
\be
h_{\rm q}=\frac12p_y^2+\frac14y^4,
\ee
the exact leading prefactor is
\be
C_{\rm c}=\frac12\langle y^2\rangle_{0}
=0.228060\ldots,
\ee
where $|0\rangle$ is the ground state of $h_{\rm q}$. The Gaussian
value is $C_{\rm c}^{(\mathrm G)}=0.218395\ldots$.
{We note that} our bounds recover the exact exponent without requiring the solution
of the critical quartic oscillator.

For every fixed $\gamma>1$, Eq.~\eqref{eq:superquad} is already an explicit error estimate {for finite $\Omega/\omega_0$}. Using Eq.~\eqref{eq:Ncl}, it reads
\begin{equation}
\left|N_g-N_{\rm cl}\right|\leq\gamma\sqrt{\frac{\eta}{2}},
\quad
\left|\frac{N_g}{\eta}-\frac{\gamma^2-\gamma^{-2}}{4}\right|
\leq\frac{\gamma}{\sqrt{2\eta}}.
\label{eq:expliciterror}
\end{equation}
Thus $N_g=N_{\rm cl}+{\cal O}(\sqrt\eta)$ with a known prefactor, not merely an unspecified asymptotic remainder. Whenever $N_{\rm cl}>0$, the relative deviation obeys
\begin{equation}
\frac{|N_g-N_{\rm cl}|}{N_{\rm cl}}
\leq
\frac{2\sqrt2\,\gamma^3}
{(\gamma^4-1)\sqrt\eta}.
\end{equation}
The full energy-derived interval has width $\Delta N=\gamma\sqrt{2\eta}$ once its lower root is positive.

The most compact consequence is obtained by introducing the intensive occupation {number}
\begin{equation}
n\equiv\frac{N_g}{\eta}
=\frac{\omega_0}{\Omega}N_g.
\label{eq:intensiveoccupation}
\end{equation}
\begin{corollary}
For every fixed $\gamma\geq0$,
\begin{equation}
\lim_{\eta\to\infty}n=
\begin{cases}
0, & \gamma\leq1,\\[1ex]
\dfrac14(\gamma^2-\gamma^{-2}), & \gamma>1.
\end{cases}
\label{eq:scaledlimit}
\end{equation}
\end{corollary}
For $\gamma<1$ this follows from Eq.~\eqref{eq:normalboundlimit}; at $\gamma=1$ the stronger result $N_g={\cal O}(\eta^{1/3})$ follows from Eq.~\eqref{eq:criticaltheta}; and for $\gamma>1$ the limit follows directly from Eq.~\eqref{eq:expliciterror}. Close to threshold, $\gamma=1+\epsilon$ with $\epsilon>0$, one has $n=\epsilon+{\cal O}(\epsilon^2)$, so the macroscopic boson fraction {starts} linearly above the transition.

\section{Numerical tests}
\label{sec:numerics}

\begin{figure}[tb]
\centering
\includegraphics[width=0.45\textwidth]{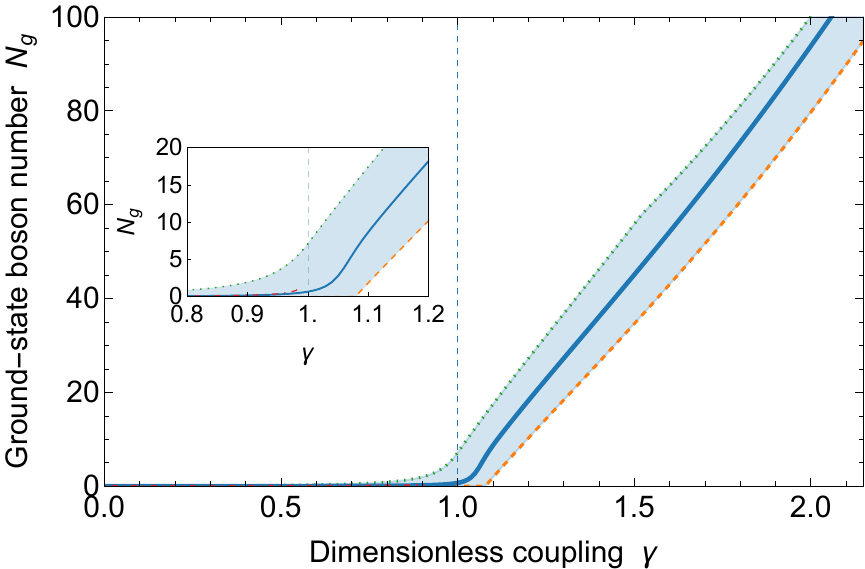}
\caption{Ground-state boson number $N_g$ versus $\gamma$ for $\eta=\Omega/\omega_0=100$. The shaded region uses the energy-derived lower edge $N_{\rm L}^{(E)}$ and the final upper bound $N_{\rm U}$. The full rigorous lower bound is $N_{\rm L}=\max\{N_{\rm L}^{(E)},N_{\rm L}^{(R)}\}$; the resolvent contribution $N_{\rm L}^{(R)}$ from Eq.~\eqref{eq:resolventlower} is positive but too small to resolve on this scale. The solid line is exact diagonalization. The dash-dotted line shows the normal-phase squeezed-oscillator asymptotics, Eq.~\eqref{eq:Nnormal}, away from the immediate critical region. The inset magnifies the crossover near $\gamma=1$.}
\label{fig:bounds}
\end{figure}

\begin{figure}[tb]
\centering
\includegraphics[width=0.45\textwidth]{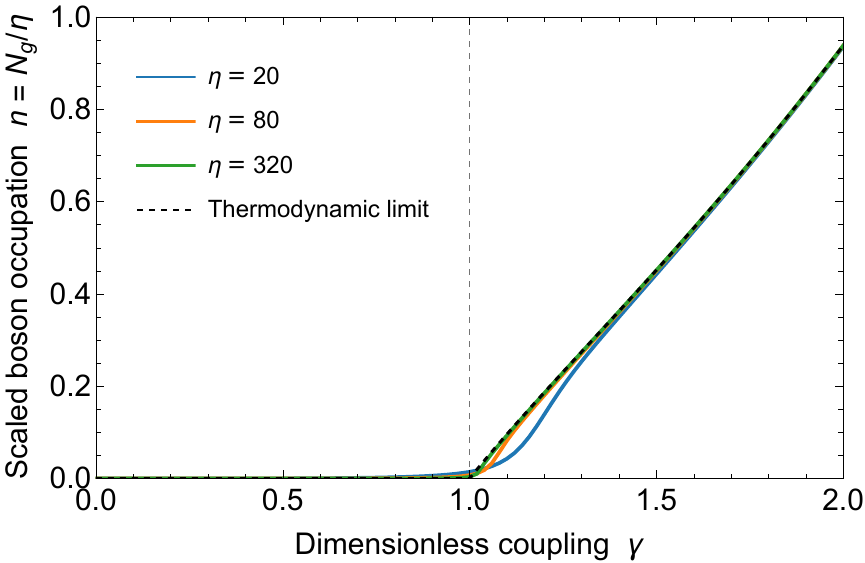}
\caption{Scaled occupation $n=(\omega_0/\Omega)N_g$ from exact diagonalization for $\eta=20$, $80$, and $320$. The dashed curve is the rigorous limiting result  Eq.~\eqref{eq:scaledlimit}. The vertical line marks $\gamma=1$.}
\label{fig:scaled}
\end{figure}

\begin{figure}[tb]
\centering
\includegraphics[width=0.45\textwidth]{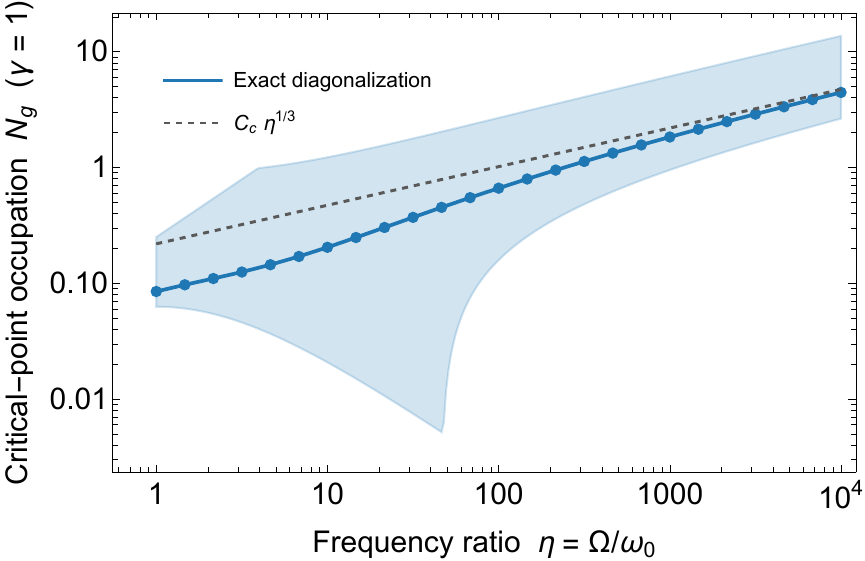}
\caption{Critical-point occupation $N_g(\gamma=1)$ versus $\eta$ on logarithmic axes. Points connected by the solid line are exact diagonalization. The dashed line is the known asymptotic result $C_{\rm c}\eta^{1/3}$ from Ref.~\cite{Hwang2015}. The shaded region is the refined rigorous interval obtained from Eqs.~\eqref{eq:criticalbound} and \eqref{eq:Beta}; its two endpoints have the same $\eta^{1/3}$ scaling.}
\label{fig:critical}
\end{figure}

We verified the analytical bounds by direct diagonalization of Eq.~\eqref{eq:Rabi} in a truncated oscillator Fock basis. We set $\omega_0=1$, so $\Omega=\eta$ and $g=\gamma\sqrt{\eta}/2$. The ground state was selected as the algebraically lowest eigenpair using a sparse Arnoldi iteration; in addition to the independent Fock-cutoff test, solver convergence was monitored through the residual norm $\|(H-E_g)|\Phi_g\rangle\|$. For Figs.~\ref{fig:bounds} and \ref{fig:scaled}, every plotted point was repeated after increasing the Fock cutoff by 70 states; the largest cutoff required was 648 states and the largest resulting change in $N_g$ was $1.9\times10^{-8}$. For the critical-scaling data in Fig.~\ref{fig:critical}, an adaptive cutoff was increased by 80 states, with a maximum change of $2.9\times10^{-9}$.

Figure~\ref{fig:bounds} compares the exact occupation {number} with the rigorous {upper and lower bounds}  for $\eta=100$. The exact result lies inside the bounds throughout the full coupling range. The inset resolves the critical region. In the normal phase the squeezed-oscillator expression \eqref{eq:Nnormal} accurately follows the exact result away from the finite-$\eta$ critical crossover, whereas in the superradiant regime the lower and upper bounds rapidly lock onto the extensive occupation.

Figure~\ref{fig:scaled} displays the scaled occupation $n=N_g/\eta$ for increasing $\eta$. The finite-frequency curves converge toward the piecewise limiting function in Eq.~\eqref{eq:scaledlimit}. The convergence is especially rapid away from $\gamma=1$, while the critical region narrows progressively as $\eta$ increases.

Figure~\ref{fig:critical} focuses on $\gamma=1$. Exact diagonalization approaches the known asymptotic law in Eq.~\eqref{eq:knowncritical} and remains inside the refined critical interval in Eq.~\eqref{eq:criticalbound}. Both rigorous endpoints scale as $\eta^{1/3}$, so the numerical data now illustrate not only the subextensivity but the correctly bounded critical exponent. The shallow nonmonotonicity of the lower envelope at intermediate $\eta$ is a consequence of taking the maximum of the competing lower estimates in Eq.~\eqref{eq:criticalbound}: the decreasing resolvent bound controls smaller $\eta$, while the quartic critical bound eventually takes over and grows as $\eta^{1/3}$.

\paragraph*{Fock-space tail estimate --}

The occupation ceiling can be converted directly into a state-space truncation criterion. Let $\Pi_{\geq M}$ project onto oscillator levels $m\geq M$, i.e. onto the part omitted when the retained Fock space contains $m=0,\ldots,M-1$. Since
\begin{equation}
a^\dagger a\geq M\Pi_{\geq M},
\label{eq:numberprojector}
\end{equation}
\begin{equation}
p_{\rm tail}(M)\equiv\langle\Phi_g|\Pi_{\geq M}|\Phi_g\rangle
\leq\frac{N_g}{M}\leq\frac{N_{\rm U}}{M}.
\label{eq:tailbound}
\end{equation}
Thus $M\geq N_{\rm U}/\varepsilon$ is a sufficient condition for the exact ground-state probability outside the retained oscillator space to be at most $\varepsilon$. This Markov-type estimate is intentionally conservative and does not by itself bound the truncated ground-state energy error, but it is fully nonperturbative and requires no knowledge of the photon-number distribution beyond the mean. In the superradiant regime, the sufficient cutoff supplied by Eq.~\eqref{eq:tailbound} therefore scales as $O(\eta)$ for any fixed target $\varepsilon$.

\section{Experimental feasibility and tests of the finite-frequency bounds}
\label{sec:experiment}

The finite-frequency bounds derived above involve the ground-state boson occupation $N_g=\langle a^\dagger a\rangle_g$, which is directly accessible in existing quantum simulators of the quantum Rabi model. In particular, trapped-ion implementations provide an attractive setting because the effective atomic and bosonic frequencies and their coupling can be tuned independently over a broad range, including the regime $\eta=\Omega/\omega_0\gg1$ relevant to the Rabi-model quantum phase transition \cite{Pedernales2015,Lv2018,Puebla2017,Cai2021}. Moreover, the bosonic degree of freedom is represented by a motional mode of the ion, whose mean phonon occupation and, when required, its Fock-state distribution can be measured experimentally. The ground-state transition has already been probed with a single trapped ion by measuring both spin and phonon populations while adiabatically varying the effective light-matter coupling \cite{Cai2021}. The present results suggest a more quantitative use of essentially the same setting: rather than identifying only the transition itself, one can directly test the finite-$\eta$ inequalities for $N_g$.

A standard trapped-ion realization uses simultaneous near-red- and near-blue-sideband drives. Denoting their detunings from the corresponding motional sidebands by $\delta_r$ and $\delta_b$, and denoting by $\Omega_R$ the balanced carrier Rabi frequency of the two drives, the interaction can be transformed to an effective quantum Rabi Hamiltonian of the form of Eq.~\eqref{eq:Rabi}. We use the sign convention in which the chosen detunings give positive effective frequencies $\Omega$ and $\omega_0$, and we choose the bichromatic phases so that the effective coupling is written along $\sigma_x$; other phase choices correspond to an equivalent rotation of the spin axis. To avoid confusion with the frequency ratio $\eta$ used throughout this work, we denote the Lamb-Dicke parameter by $\eta_{\rm LD}$. With these conventions, the correspondence between the simulated and experimental parameters is
\begin{equation}
\Omega=\frac{\delta_b+\delta_r}{2},
\quad
\omega_0=\frac{\delta_b-\delta_r}{2},
\quad
g=\frac{\eta_{\rm LD}\Omega_R}{2}.
\label{eq:ionmapping}
\end{equation}
Consequently,
\begin{equation}
\eta=\frac{\Omega}{\omega_0}
=\frac{\delta_b+\delta_r}{\delta_b-\delta_r},
\quad
\gamma=
\frac{2\eta_{\rm LD}\Omega_R}
{\sqrt{\delta_b^2-\delta_r^2}}.
\label{eq:iondimensionless}
\end{equation}
Thus $\eta$ can be varied through the bichromatic detunings while $\gamma$ can be scanned through the sideband-drive strength. This tunability is particularly useful here because the theoretical bounds apply for arbitrary finite values of both parameters.

A direct test of Eq.~\eqref{eq:finalbounds} can be performed by preparing approximate ground states at a sequence of points $(\eta,\gamma)$ and measuring the mean phonon occupation,
\begin{equation}
N_{\rm exp}=\sum_{m=0}^{\infty}m p_m,
\label{eq:Nexp}
\end{equation}
where $p_m$ is the measured population of motional Fock state $|m\rangle$. Starting from a sideband-cooled motional state and the spin ground state, the effective coupling may be ramped adiabatically from $\gamma=0$ to its desired final value. Repeating the measurement as a function of $\gamma$ at fixed $\eta$ would provide an experimental analogue of Fig.~\ref{fig:bounds}. For a sufficiently accurate realization of the QRM ground state, the resulting points should obey
\begin{equation}
N_{\rm L}(\eta,\gamma)
\leq N_{\rm exp}(\eta,\gamma)
\leq N_{\rm U}(\eta,\gamma),
\label{eq:experimentalinterval}
\end{equation}
within experimental uncertainties.

The three parameter regimes provide qualitatively different experimental tests. For any fixed $\gamma<1$, the occupation remains $O(1)$ as $\eta$ is increased. An experiment performed at several fixed subcritical couplings would therefore verify the absence of extensive bosonic occupation and could additionally compare with the squeezed-oscillator expression in Eq.~\eqref{eq:Nnormal}. At the critical coupling $\gamma=1$, the stronger prediction is
\begin{equation}
N_g={\cal O}(\eta^{1/3}).
\label{eq:experimentalcritical}
\end{equation}
This can be tested by varying the frequency ratio $\eta$, preparing the state at $\gamma=1$, and plotting $N_{\rm exp}$ against $\eta$ on logarithmic axes. The measured data should remain between the two rigorous curves of Eq.~\eqref{eq:criticalbound}, while their asymptotic logarithmic slope should approach $1/3$. Such a measurement would go beyond detecting the critical point: it would test directly the finite-frequency critical interval obtained here.

The superradiant regime provides an especially simple quantitative test. For fixed $\gamma>1$, Eq.~\eqref{eq:expliciterror} predicts
\begin{equation}
\left|
N_g-
\frac{\eta}{4}
(\gamma^2-\gamma^{-2})
\right|
\leq
\gamma\sqrt{\frac{\eta}{2}}.
\label{eq:experimental_super_bound}
\end{equation}
It is useful to introduce the normalized deviation
\begin{equation}
D(\eta,\gamma)
=
\frac{
\left|N_{\rm exp}-N_{\rm cl}\right|
}{
\gamma\sqrt{\eta/2}
},
\quad
N_{\rm cl}=
\frac{\eta}{4}(\gamma^2-\gamma^{-2}).
\label{eq:normalized_deviation}
\end{equation}
For an accurately prepared QRM ground state the theorem requires
\begin{equation}
D(\eta,\gamma)\leq1.
\label{eq:Dbound}
\end{equation}
Measurements for several values of $\eta$ at fixed $\gamma>1$ would therefore provide a dimensionless test of the finite-frequency error estimate. In addition, plotting $N_{\rm exp}/\eta$ would directly display its convergence toward $(\gamma^2-\gamma^{-2})/4$ as the effective thermodynamic limit is approached.

The Fock-space tail estimate in Eq.~\eqref{eq:tailbound} also admits a direct experimental interpretation. If the motional distribution $p_m$ is reconstructed rather than only its mean being measured, one can form
\begin{equation}
P_{\geq M}^{\rm exp}
=
\sum_{m=M}^{\infty}p_m .
\label{eq:experimental_tail}
\end{equation}
The theoretical result predicts
\begin{equation}
P_{\geq M}^{\rm exp}
\leq
\frac{N_g}{M}
\leq
\frac{N_{\rm U}}{M},
\label{eq:experimental_tail_bound}
\end{equation}
again provided that the prepared state approximates the ground state of Eq.~\eqref{eq:Rabi}. Since the estimate is intentionally conservative, its experimental value would not be to determine the detailed shape of the phonon distribution, but rather to verify a parameter-only ceiling on its high-occupation tail.

The critical regime is likely to impose the strongest state-preparation requirement. The excitation gap decreases near $\gamma=1$ as the singular limit is approached, so increasing $\eta$ simultaneously improves access to universal critical scaling and makes adiabatic preparation more demanding \cite{Puebla2017,Cai2021}. A useful protocol for testing Eq.~\eqref{eq:criticalbound} is therefore to approach $\gamma=1$ from the normal side and terminate the ramp at the critical point, rather than crossing into the superradiant regime. A locally adiabatic schedule, with the ramp slowed only in the vicinity of the minimum gap, can further reduce the required preparation time. Experimentally, convergence of $N_{\rm exp}$ under increasing ramp duration provides an essential check that residual excitations do not dominate the measured occupation.

Several additional consistency tests are important because Eqs.~\eqref{eq:finalbounds} and \eqref{eq:criticalbound} are statements about the exact ground state of the ideal QRM. The effective parameters $\omega_0$, $\Omega$, and $g$ should therefore be independently calibrated, and the balance between red- and blue-sideband coupling strengths should be verified. Measurements should also be repeated for increasing ramp times and, where possible, for different motional coherence conditions. Corrections associated with the Lamb-Dicke approximation, off-resonant sidebands, drive imbalance, motional heating, and dephasing provide systematic departures from the Hamiltonian in Eq.~\eqref{eq:Rabi}. Consequently, an experimental point outside the rigorous interval would not by itself contradict the mathematical bound; it would instead signal imperfect ground-state preparation, imperfect realization of the assumed Hamiltonian, or underestimated experimental uncertainties. Conversely, observing systematic convergence into the predicted interval as these imperfections are reduced would provide a particularly stringent validation of the finite-frequency theory.

Finally, the parity discussion above suggests a complementary characterization of the superradiant regime. At any finite $\eta$, the exact ground state has definite parity and hence $\langle a\rangle_g=0$, even though $N_g$ becomes $O(\eta)$. Measurements sensitive to motional quadratures could therefore distinguish the growth of the displacement scale from an actual finite-$\eta$ symmetry-breaking expectation value: one expects $\langle x\rangle_g=0$ while $\langle x^2\rangle_g$ grows strongly above threshold. A reconstruction of the motional phase-space distribution would provide a direct visualization of the two parity-related semiclassical branches underlying the extensive occupation.

These considerations show that the finite-frequency results derived here are not restricted to an asymptotic theoretical setting. Existing trapped-ion quantum simulators already provide independent control of the parameters entering $\eta$ and $\gamma$ and direct access to the boson occupation used throughout this work. Particularly natural experimental targets are therefore the full finite-$\eta$ interval in Eq.~\eqref{eq:finalbounds}, the $\eta^{1/3}$ critical interval in Eq.~\eqref{eq:criticalbound}, the normalized superradiant constraint in Eq.~\eqref{eq:Dbound}, and the Fock-space tail estimate in Eq.~\eqref{eq:experimental_tail_bound}.

\section{Discussion and conclusions}
\label{sec:conclusions}

The three regimes are now controlled at different, but physically natural, scales. In the normal phase the bounds prove $N_g=O(1)$, although they do not fix the exact squeezed-vacuum value. At $\gamma=1$ the quadratic confinement disappears; retaining the quartic soft mode closes the previous gap and gives the two-sided result $N_g={\cal O}(\eta^{1/3})$. Above threshold the lower and upper roots share the same ${\cal O}(\eta)$ term and {approach} the thermodynamic occupation with a relative error ${\cal O}(\eta^{-1/2})$. Thus the same bounding strategy distinguishes the finite, critical, and extensive occupation scales across the transition.

The comparison with earlier work is particularly transparent in the superradiant regime. Ref.~\cite{Hwang2015} supplies the limiting center $N_{\rm cl}/\eta=(\gamma^2-\gamma^{-2})/4$, whereas Eq.~\eqref{eq:expliciterror} controls the distance of the finite-$\eta$ quantum result from that center. Ref.~\cite{HirokawaMollerSasaki2017}, on the other hand, develops rigorous dressed-photon bounds and the pull-through representation in a generalized Rabi model, with emphasis on strong-coupling photon dressing and the $A^2$ term. The present energy-derived interval addresses a different asymptotic question: fixed $\gamma>1$ with increasing $\eta$, where both endpoints share the correct extensive term and their relative separation vanishes as $\eta^{-1/2}$.

The resolvent and energy-derived bounds have complementary origins. Equation~\eqref{eq:resolventbound} follows only from the spectral gap $H-E_g+\omega_0\geq\omega_0$, while Eq.~\eqref{eq:resolventlower} follows from convexity of the inverse-square resolvent and the bounded spin polarization. The latter is asymptotically saturated in weak coupling. Equations~\eqref{eq:NL} and \eqref{eq:NUE} instead combine the coherent-state energy bound with the operator inequality \eqref{eq:operatorineq}; above threshold they become decisive because both roots are centered on the semiclassical occupation $N_{\rm cl}=|\alpha_0|^2$. Taking the maximum of the lower bounds and the minimum of the upper bounds improves the finite-parameter interval without changing the universal large-$\eta$ conclusion.

Parity is also important for interpreting the result. At finite $\eta$ the exact ground state has $\langle a\rangle_g=0$, so $N_{\rm cl}$ should not be read as the square of an exact symmetry-breaking expectation value. Rather, it is the occupation scale associated with the two parity-related semiclassical minima, and Eq.~\eqref{eq:expliciterror} proves that the exact parity-even boson number is confined to a relatively shrinking neighborhood of this scale.

Finally, Eq.~\eqref{eq:tailbound} turns the mean-occupation ceiling into a rigorous statement about Hilbert-space truncation. Although the resulting Markov bound is conservative and does not replace an explicit convergence test, it gives a parameter-only sufficient cutoff for controlling the omitted ground-state probability. This makes the analytical bounds directly useful in the strongly displaced regime, where brute-force cutoff selection becomes increasingly costly.

In summary, the bounds resolve the three characteristic occupation scales of the Rabi transition. For fixed $\gamma<1$ they prove $N_g=O(1)$. At $\gamma=1$, the quartic variational estimate together with the operator and uncertainty bounds gives $N_g=\Theta(\eta^{1/3})$, recovering the known critical exponent by a direct two-sided inequality. For fixed $\gamma>1$, Eq.~\eqref{eq:expliciterror} confines the exact boson number to an interval centered on the known thermodynamic occupation $N_{\rm cl}=|\alpha_0|^2$, with absolute width $O(\sqrt\eta)$ and an explicit $O(\eta^{-1/2})$ convergence rate for the scaled occupation. The same framework also supplies the coherent-state energy bracket of width $\omega_0$, the weak-coupling lower estimate in Eq.~\eqref{eq:resolventlower}, and the Fock-space tail bound in Eq.~\eqref{eq:tailbound}. The advance is therefore a controlled finite-frequency description across all three regimes, rather than a new thermodynamic order parameter.

\acknowledgments

NVV acknowledges support by the EU Staff Exchanges project EQSTASE.
{MF acknowledges financial support from the DFG through SFB TR 185, Project No. 277625399.}

\bibliography{rabi_references}

@article{Rabi1936,
  author  = {Rabi, I. I.},
  title   = {On the Process of Space Quantization},
  journal = {Physical Review},
  volume  = {49},
  number  = {4},
  pages   = {324--328},
  year    = {1936},
  doi     = {10.1103/PhysRev.49.324}
}

@article{Rabi1937,
  author  = {Rabi, I. I.},
  title   = {Space Quantization in a Gyrating Magnetic Field},
  journal = {Physical Review},
  volume  = {51},
  number  = {8},
  pages   = {652--654},
  year    = {1937},
  doi     = {10.1103/PhysRev.51.652}
}

@article{JaynesCummings1963,
  author  = {Jaynes, E. T. and Cummings, F. W.},
  title   = {Comparison of Quantum and Semiclassical Radiation Theories with Application to the Beam Maser},
  journal = {Proceedings of the IEEE},
  volume  = {51},
  number  = {1},
  pages   = {89--109},
  year    = {1963},
  doi     = {10.1109/PROC.1963.1664}
}

@article{EberlyNarozhnySanchezMondragon1980,
  author  = {Eberly, Joseph H. and Narozhny, N. B. and Sanchez-Mondragon, J. J.},
  title   = {Periodic Spontaneous Collapse and Revival in a Simple Quantum Model},
  journal = {Physical Review Letters},
  volume  = {44},
  number  = {20},
  pages   = {1323--1326},
  year    = {1980},
  doi     = {10.1103/PhysRevLett.44.1323}
}

@article{NarozhnySanchezMondragonEberly1981,
  author  = {Narozhny, N. B. and Sanchez-Mondragon, J. J. and Eberly, Joseph H.},
  title   = {Coherence versus Incoherence: Collapse and Revival in a Simple Quantum Model},
  journal = {Physical Review A},
  volume  = {23},
  number  = {1},
  pages   = {236--247},
  year    = {1981},
  doi     = {10.1103/PhysRevA.23.236}
}

@article{AgarwalRafsanjaniEberly2013,
  author  = {Agarwal, S. and Hashemi Rafsanjani, S. M. and Eberly, Joseph H.},
  title   = {Dissipation of the {Rabi} Model beyond the {Jaynes--Cummings} Approximation: Quasi-Degenerate Qubit and Ultra-Strong Coupling},
  journal = {Journal of Physics B: Atomic, Molecular and Optical Physics},
  volume  = {46},
  number  = {22},
  pages   = {224017},
  year    = {2013},
  doi     = {10.1088/0953-4075/46/22/224017}
}

@article{Braak2011,
  author  = {Braak, Daniel},
  title   = {Integrability of the {Rabi} Model},
  journal = {Physical Review Letters},
  volume  = {107},
  number  = {10},
  pages   = {100401},
  year    = {2011},
  doi     = {10.1103/PhysRevLett.107.100401}
}

@article{Chen2012,
  author  = {Chen, Qing-Hu and Wang, Chen and He, Shu and Liu, Tao and Wang, Ke-Lin},
  title   = {Exact Solvability of the Quantum {Rabi} Model Using {Bogoliubov} Operators},
  journal = {Physical Review A},
  volume  = {86},
  number  = {2},
  pages   = {023822},
  year    = {2012},
  doi     = {10.1103/PhysRevA.86.023822}
}

@article{Xie2017,
  author  = {Xie, Qiongtao and Zhong, Honghua and Batchelor, Murray T. and Lee, Chaohong},
  title   = {The Quantum {Rabi} Model: Solution and Dynamics},
  journal = {Journal of Physics A: Mathematical and Theoretical},
  volume  = {50},
  number  = {11},
  pages   = {113001},
  year    = {2017},
  doi     = {10.1088/1751-8121/aa5a65}
}

@article{Niemczyk2010,
  author  = {Niemczyk, T. and Deppe, F. and Huebl, H. and Menzel, E. P. and Hocke, F. and Schwarz, M. J. and Garcia-Ripoll, J. J. and Zueco, D. and Hummer, T. and Solano, E. and Marx, A. and Gross, R.},
  title   = {Circuit Quantum Electrodynamics in the Ultrastrong-Coupling Regime},
  journal = {Nature Physics},
  volume  = {6},
  pages   = {772--776},
  year    = {2010},
  doi     = {10.1038/nphys1730}
}

@article{Casanova2010,
  author  = {Casanova, J. and Romero, G. and Lizuain, I. and Garcia-Ripoll, J. J. and Solano, E.},
  title   = {Deep Strong Coupling Regime of the {Jaynes--Cummings} Model},
  journal = {Physical Review Letters},
  volume  = {105},
  number  = {26},
  pages   = {263603},
  year    = {2010},
  doi     = {10.1103/PhysRevLett.105.263603}
}

@article{Yoshihara2017,
  author  = {Yoshihara, Fumiki and Fuse, Tomoko and Ashhab, Sahel and Kakuyanagi, Kosuke and Saito, Shiro and Semba, Kouichi},
  title   = {Superconducting Qubit--Oscillator Circuit beyond the Ultrastrong-Coupling Regime},
  journal = {Nature Physics},
  volume  = {13},
  number  = {1},
  pages   = {44--47},
  year    = {2017},
  doi     = {10.1038/nphys3906}
}

@article{FornDiaz2019,
  author  = {Forn-D{\'i}az, P. and Lamata, L. and Rico, E. and Kono, J. and Solano, E.},
  title   = {Ultrastrong Coupling Regimes of Light-Matter Interaction},
  journal = {Reviews of Modern Physics},
  volume  = {91},
  number  = {2},
  pages   = {025005},
  year    = {2019},
  doi     = {10.1103/RevModPhys.91.025005}
}

@article{Pedernales2015,
  author  = {Pedernales, J. S. and Lizuain, I. and Felicetti, S. and Romero, G. and Lamata, L. and Solano, E.},
  title   = {Quantum {Rabi} Model with Trapped Ions},
  journal = {Scientific Reports},
  volume  = {5},
  pages   = {15472},
  year    = {2015},
  doi     = {10.1038/srep15472}
}

@article{Lv2018,
  author  = {Lv, Dingshun and An, Shuoming and Liu, Zhenyu and Zhang, Jing-Ning and Pedernales, Julen S. and Lamata, Lucas and Solano, Enrique and Kim, Kihwan},
  title   = {Quantum Simulation of the Quantum {Rabi} Model in a Trapped Ion},
  journal = {Physical Review X},
  volume  = {8},
  number  = {2},
  pages   = {021027},
  year    = {2018},
  doi     = {10.1103/PhysRevX.8.021027}
}

@article{Ashhab2013,
  author  = {Ashhab, S.},
  title   = {Superradiance Transition in a System with a Single Qubit and a Single Oscillator},
  journal = {Physical Review A},
  volume  = {87},
  number  = {1},
  pages   = {013826},
  year    = {2013},
  doi     = {10.1103/PhysRevA.87.013826}
}

@article{Puebla2017,
  author  = {Puebla, Ricardo and Hwang, Myung-Joong and Casanova, Jorge and Plenio, Martin B.},
  title   = {Probing the Dynamics of a Superradiant Quantum Phase Transition with a Single Trapped Ion},
  journal = {Physical Review Letters},
  volume  = {118},
  number  = {7},
  pages   = {073001},
  year    = {2017},
  doi     = {10.1103/PhysRevLett.118.073001}
}

@article{Hwang2015,
  author  = {Hwang, Myung-Joong and Puebla, Ricardo and Plenio, Martin B.},
  title   = {Quantum Phase Transition and Universal Dynamics in the {Rabi} Model},
  journal = {Physical Review Letters},
  volume  = {115},
  number  = {18},
  pages   = {180404},
  year    = {2015},
  doi     = {10.1103/PhysRevLett.115.180404}
}

@article{Puebla2016,
  author  = {Puebla, Ricardo and Hwang, Myung-Joong and Plenio, Martin B.},
  title   = {Excited-State Quantum Phase Transition in the {Rabi} Model},
  journal = {Physical Review A},
  volume  = {94},
  number  = {2},
  pages   = {023835},
  year    = {2016},
  doi     = {10.1103/PhysRevA.94.023835}
}

@article{Cai2021,
  author  = {Cai, M.-L. and Liu, Z.-D. and Zhao, W.-D. and Wu, Y.-K. and Mei, Q.-X. and Jiang, Y. and He, L. and Zhang, X. and Zhou, Z.-C. and Duan, L.-M.},
  title   = {Observation of a Quantum Phase Transition in the Quantum {Rabi} Model with a Single Trapped Ion},
  journal = {Nature Communications},
  volume  = {12},
  pages   = {1126},
  year    = {2021},
  doi     = {10.1038/s41467-021-21425-8}
}

@article{Liu2015,
  author  = {Liu, Maoxin and Ying, Zu-Jian and An, Jun-Hong and Luo, Hong-Gang},
  title   = {Mean Photon Number Dependent Variational Method to the {Rabi} Model},
  journal = {New Journal of Physics},
  volume  = {17},
  number  = {4},
  pages   = {043001},
  year    = {2015},
  doi     = {10.1088/1367-2630/17/4/043001}
}

@article{Ying2015,
  author  = {Ying, Zu-Jian and Liu, Maoxin and Luo, Hong-Gang and Lin, Hai-Qing and You, J. Q.},
  title   = {Ground-State Phase Diagram of the Quantum {Rabi} Model},
  journal = {Physical Review A},
  volume  = {92},
  number  = {5},
  pages   = {053823},
  year    = {2015},
  doi     = {10.1103/PhysRevA.92.053823}
}

@article{Gartner2022,
  author  = {Gartner, P. and Moldoveanu, V.},
  title   = {Effective Hamiltonians in the Quantum {Rabi} Problem},
  journal = {Physical Review A},
  volume  = {105},
  number  = {2},
  pages   = {023704},
  year    = {2022},
  doi     = {10.1103/PhysRevA.105.023704}
}

@article{AraiHirokawa1997,
  author  = {Arai, Asao and Hirokawa, Masao},
  title   = {On the Existence and Uniqueness of Ground States of a Generalized Spin-Boson Model},
  journal = {Journal of Functional Analysis},
  volume  = {151},
  number  = {2},
  pages   = {455--503},
  year    = {1997},
  doi     = {10.1006/jfan.1997.3140}
}

@article{HirokawaMollerSasaki2017,
  author  = {Hirokawa, Masao and M{\o}ller, Jacob S. and Sasaki, Itaru},
  title   = {A Mathematical Analysis of Dressed Photon in Ground State of Generalized Quantum {Rabi} Model Using Pair Theory},
  journal = {Journal of Physics A: Mathematical and Theoretical},
  volume  = {50},
  number  = {18},
  pages   = {184003},
  year    = {2017},
  doi     = {10.1088/1751-8121/aa677c}
}

@article{Lieb1973,
  author  = {Lieb, Elliott H.},
  title   = {The Classical Limit of Quantum Spin Systems},
  journal = {Communications in Mathematical Physics},
  volume  = {31},
  number  = {4},
  pages   = {327--340},
  year    = {1973},
  doi     = {10.1007/BF01646493}
}

@article{Berezin1975,
  author  = {Berezin, F. A.},
  title   = {General Concept of Quantization},
  journal = {Communications in Mathematical Physics},
  volume  = {40},
  number  = {2},
  pages   = {153--174},
  year    = {1975}
}

@article{Hansen2003,
  author  = {Hansen, Frank and Pedersen, Gert K.},
  title   = {Jensen's Operator Inequality},
  journal = {Bulletin of the London Mathematical Society},
  volume  = {35},
  number  = {4},
  pages   = {553--564},
  year    = {2003},
  doi     = {10.1112/S0024609303002200}
}

@article{dimer2007proposed,
  title={Proposed realization of the Dicke-model quantum phase transition in an optical cavity QED system},
  author={Dimer, F and Estienne, B and Parkins, AS and Carmichael, HJ},
  journal={Physical Review A—Atomic, Molecular, and Optical Physics},
  volume={75},
  number={1},
  pages={013804},
  year={2007},
  publisher={APS}
}

@article{weitz2023,
  author = {Koch, Johannes and
            Hunanyan, Geram R. and
            Ockenfels, Till and
            Rico, Enrique and
            Solano, Enrique and
            Weitz, Martin},
  title = {Quantum Rabi dynamics of trapped atoms
           far in the deep strong coupling regime},
  journal = {Nature Communications},
  volume = {14},
  pages = {954},
  year = {2023},
  doi = {10.1038/s41467-023-36611-z}
}

@article{Weitz2024,
  author = {Hunanyan, Geram R. and
            Koch, Johannes and
            Moll, Stefanie and
            Rico, Enrique and
            Solano, Enrique and
            Weitz, Martin},
  title = {Periodic quantum Rabi model with
           cold atoms at deep strong coupling},
  journal = {Physical Review Research},
  volume = {6},
  pages = {033023},
  year = {2024},
  doi = {10.1103/PhysRevResearch.6.033023}
}

\end{document}